\documentclass[acmcsur]{acmsmall} 
\DeclareRobustCommand{\disambiguate}[3]{#3}

\acmYear{2014}
\usepackage{color}
\usepackage{url}
\def\UrlBreaks{\do\/\do-}
\begin{document}

\markboth{S. Mittal and J. S. Vetter}{A Survey of Methods For Analyzing and Improving GPU Energy Efficiency}

\title{A Survey of Methods For Analyzing and Improving GPU Energy Efficiency}
\author{Sparsh Mittal
\affil{Iowa State University}
Jeffrey S. Vetter
\affil{Oak Ridge National Laboratory}
}

\begin{abstract}

Recent years have witnessed a phenomenal growth in the computational capabilities and applications of GPUs. However, this trend has also led to dramatic increase in their power consumption. This paper surveys research works on analyzing and improving energy efficiency of GPUs. It also provides a classification of these techniques on the basis of their main research idea. Further, it attempts to synthesize research works which compare energy efficiency of GPUs with other computing systems, e.g. FPGAs and CPUs. The aim of this survey is to provide researchers with knowledge of state-of-the-art in GPU power management and motivate them to architect highly energy-efficient GPUs of tomorrow.

\end{abstract}

 

            

\acmformat{Sparsh Mittal and Jeffrey S. Vetter, 2014. A Survey of Methods For Analyzing and Improving GPU Energy Efficiency. Accepted with minor revision in}

\begin{bottomstuff}
Authors' address: Sparsh Mittal and Jeffrey S. Vetter, 1 Bethel Valley Road, Future Technologies Group, Oak Ridge National Laboratory, Building 5100,  MS-6173, Tennessee USA 37830; email: \{mittals,vetter\}@ornl.gov
\end{bottomstuff}

\maketitle

\section{Introduction}
As we enter into the post-petascale era, the requirements of data processing and computation are growing exponentially. To meet this requirement, researchers have moved from serial execution platforms to high-performance  computing (HPC) platforms, such as multicore processors, FPGAs and GPUs etc. GPUs, in particular, have been widely used for HPC applications due to their extremely high computational powers, and a large fraction of supercomputers in Top500 list use GPU to achieve unprecedented computational power \cite{top500listlatest}. Thus, GPUs have become integral part of today's mainstream computing systems.  

The high performance demands on GPUs, however, have influenced their design to be  optimized for higher performance, even at the cost of large power consumption. Hence, recent years have witnessed marked increase in power consumption of GPUs.  
The elevated levels of power consumption of GPUs have significant impact on their reliability, economic feasibility, architecture design, performance scaling and deployment into a wide range of application domains. As a case in point, supercomputers built with CPU-GPU  consume huge amount of power, for example, Titan supercomputer consumes 8.2MW power \cite{top500listlatest}. 
 Further, it has been estimated that an exascale machine, built with the technology used in today's supercomputers will consume several giga watts of power  \cite{exascaleMachineChallenge}.  To manage such high levels of power dissipation and continue to scale performance, power management techniques are essential for both CPUs and GPUs. While the area of power management in CPUs has been actively researched over years, the area of power management in GPUs is yet to be fully explored. For these reasons, understanding the state-of-the-art in GPU power management in extremely important for researchers to propose even more effective solutions to address the power challenges and design ``green'' GPUs of tomorrow.

In this paper, we present a survey of research works aimed at analyzing and improving energy efficiency of GPUs. We classify the techniques based on several parameters to provide insights into their important features. We also review the research works which compare the energy efficiency of GPUs with other computing systems such as CPUs, Cell processor, FPGA etc. We believe that this will enable the readers to judge the  energy efficiency of GPUs vis-\`a-vis alternate computing platforms and make important decisions.
 
Since it is infeasible to review all the research ideas proposed in the literature, we adopt the following approach to limit the scope of the paper. We include only those studies that analyze GPU power consumption and the techniques which have been evaluated based on GPU energy efficiency. We do not include those studies which have been shown to improve only performance and not energy efficiency, even though the performance improvement is likely to translate to better energy efficiency. We include application-level and architectural-level techniques and not circuit-level techniques  for improving energy efficiency. Further, since different techniques have been evaluated using different experimentation platform and methodologies, we only focus on their key ideas and generally do not present their quantitative results.

This paper is organized as follows. Section \ref{sec:background} reviews the GPU terminology and also highlights the need of power management. Section \ref{sec:analysis} reviews the studies on comparing GPU energy efficiency with that of other computing systems.  Section \ref{sec:improvement} discusses some power management techniques in detail. In both of these sections, we first provide an overview and classification of the methods; and then discuss some of the techniques in detail. We finally  provide concluding remarks and future research trends in Section \ref{sec:conclusion}. 

\section{Background} \label{sec:background}

\subsection{ GPU Terminology and Sources of Power Consumption}
Recently, several researchers have proposed models and tools for measurement and estimation of GPU power consumption \cite{hong2010integrated,ramani2007powerred,nagasaka2010statistical,sheaffer2005fine,zhang2011performance,jiao2010power,zhang2011architecture,chen2011tree,suda2009accurate,enos2010quantifying,6036835,ren2011algorithm,6136741,luo2011performance,pool2010energy,stolz2010energy,li2011energy,haifeng2012,collange2009power,wang2010analysis,vialle2011optimizing,kasichayanula2012}. These models provide insights into the working of GPUs and relative contribution of different components in the total power consumption. In what follows, we briefly review the GPU architecture, terminology and sources of power consumption, as relevant for this paper  and refer the reader to above mentioned works for more details.

A GPU has several streaming multiprocessors, each of which has multiple cores. For example, NVIDIA GeForce GTX 590 has dual GPUs; where each GPU has 16 streaming multiprocessors (SMs); each of these SMs have 32 cores; for a total of 512 cores in each GPU and 1024 cores in the overall GTX 590 graphics card \cite{GPU590power}.  
The cores of a typical GPU are composed of ALUs, thread-schedulers, load/store units,
scratchpad memory, register file and caches etc. A GPU is designed for stream or throughput computing, which has little data reuse and hence, a GPU has much smaller sized cache (for example 16KB L1 and 256KB L2 \cite{wong2010demystifying}) than a typical CPU. The GPU is used as a co-processor with a CPU and in such cases, GPU is referred to as the `device' and the CPU as the `host'. A GPU has its own device memory of a few GBs (gigabytes), and it is connected to the host through a PCI-Express (PCIe) bus. A GPU is programmed as a sequence of \textit{kernels}. The code is  executed in groups of 32 threads, called a \textit{warp}. CUDA (Compute Unified Device Architecture) and OpenCL (Open Computing Language) are widely-used interfaces for programming GPUs.    

The power consumption of GPU can be divided into two parts, namely leakage power and dynamic power. The dynamic power is a function of operating temperature and circuit technology. Leakage power is consumed when GPU is powered, even if there are no runtime activities. The dynamic power arises from switching of transistors and is determined by the runtime activities. Different components such as SMs and memories (e.g local, global, shared) etc. contribute to this power consumption.

\subsection{Need for Improving Energy Efficiency of GPUs}
GPU power management is extremely important for the following reasons. 
\subsubsection{Addressing Inefficient Resource Usage}

To meet the worst-case performance requirements, the chip  designers need to over-provision the computing resources of GPUs; however, on average, the utilization of these resources remains low. Also, in several applications, memory bandwidth of GPUs acts as a performance-bottleneck \cite{hong2010integrated,daga2011efficacy,6270749,spafford2012tradeoffs}, due to which the cores are not fully utilized which leads to energy inefficiency. Further, unlike massively parallel applications, regular parallel applications do not scale well beyond a certain number of cores and hence, a large amount of power is wasted in idle cores or in synchronization. Finally, GPUs are increasingly being used in cloud infrastructure and data centers \cite{amazoncloud}, which experience highly varying usage patterns. Thus, dynamic power management techniques can offset these sources of inefficiencies by using runtime adaption. 

\subsubsection{Ensuring Reliability}
Large power consumption has significant effect on the reliability of the computing systems. A 15 degree Celsius rise in temperature increases the component failure rates by up to a factor of two \cite{anderson2003more}. The device failures may lead to system malfunction and as GPUs become increasingly employed in supercomputers  and business services, system malfunction may have serious economic impact. For example, the service cost of merely one hour of downtime in brokerage operations and credit card authorization can be \$6,450,000 and \$2,600,000, respectively \cite{feng2003making}. Thus, since the performance-requirements grow at much faster pace than the effectiveness of cooling solutions, power management techniques are  extremely important to ensure longevity and reliability. 

\subsubsection{Providing Economic Gains}
For every watt of power dissipated in the computing equipment, an additional 0.5 to 1W of power is consumed by the cooling system itself \cite{patel2003smart}, and with increasing ratio of cooling power to computing power, compaction of devices is inhibited, which results in increased operation costs. Due to these trends, in recent years, the energy cost of high-performance computing clusters has been estimated to contribute more than the hardware acquisition cost of IT equipment itself \cite{bianchini2004power,Mit_DRAMsurvey}.

\subsubsection{Enabling Performance Scaling}
The power challenges are expected to present most severe obstacle to performance scaling and it has been shown that thermal and leakage power constraints may disallow simultaneously using all the cores of a massively parallel processor \cite{esmaeilzadeh2013power}. Large power consumption may necessitate complex cooling solutions (e.g. liquid cooling) which may increase chip complexity and  offset the benefits of performance boost obtained by using GPUs.

\subsubsection{Enabling Deployment in Wide Range of Applications}

The energy efficiency of GPUs, relative to other alternatives (e.g. CPUs, FPGAs) will  have a crucial role in deciding its adoption in various application domains. In recent years, ongoing technological innovations have greatly improved other computing systems. As we show in Section  \ref{sec:analysis}, for several applications FPGAs have been found to have better performance and energy efficiency than GPUs. Moreover, while a few initial works have reported orders of magnitude difference in performance of GPUs and CPUs, other researchers who apply careful optimization on \textit{both} CPUs and GPUs have reported much lower speedups of GPUs over CPUs, typically in the range of $0.7\times$ to $15\times$ \cite{lee2010debunking,zou2012optimization,chandramowlishwaran2010optimizing}. Thus, to maintain  their competitiveness and justify their use in product design, GPUs must exhibit high energy efficiency.  

\subsubsection{Achieving the Goals of Sustainable Computing}
It has been estimated that the carbon emission of ICT (information and communication technology) will triple from 2002 to 2020 \cite{smarr2010project} and hence, the concerns for environment will force the policy-makers and researchers to place higher emphasis on energy efficiency in the design of future computing systems. Thus, improving the energy efficiency of GPUs is also important for achieving the goals of sustainable computing.

\section{Research Works on Analyzing GPU Energy Efficiency}\label{sec:analysis}
In this section, we review the  research works which analyze energy efficiency of GPUs and compare it with that of other computing systems. We first present an overview and then discuss some of the research works in detail.  
\subsection{Overview}
Modern GPUs consume significant amount of power. The high-end GPUs, such as NVIDIA GeForce GTX 590 (40nm) and  AMD Radeon HD 5970 (40nm) have a  maximum power consumption of 365W \cite{GPU590power}, and 294W \cite{AMD5970power}, respectively. In contrast, Intel's Core i7-3770T (22nm), Xeon E7-8870 (32nm) and have a maximum power consumption of 45W and 150W, respectively \cite{intelpowerwww,intelpowerwww3}. Note, however, that for several applications, GPUs provide better performance than CPUs which makes their energy efficiency better than those of CPUs. 

In recent years, several researchers have compared the power consumption of GPUs with that of other computing systems such as CPUs, Cell or FPGA. For certain applications and platforms GPUs have been found to be more energy efficient than CPUs \cite{zandevakili2012gpumotif,huang2009energy,anzt2011analysis,baker2007matched,thomas2009comparison,hamada2009comparative,mcintosh2012benchmarking,lange2009acceleration,ghosh2012energy,udagawa2011power,zou2012optimization,hussain2011highly,de2011energy,van2012accelerating,betkaoui2010comparing,timm2010reducing,goddeke2008using,scogland2010first,danalis2010scalable,chung2010single,keckler2011gpus,brodtkorb2010state,thomas2013gpufpga,lopez2011gpgpu,cong20113d,pedram2012co,chow2012mixed,wang2012coarse}, while for other applications CPUs have been found to be more energy efficient \cite{chandramowlishwaran2010optimizing,kestur2010blas}. Some researchers also discuss the conditions under which CPUs or GPUs may be more efficient \cite{datta2008stencil,anzt2010energy,calandrini2012gpu,fowers2013performance,Maghazeh13}. For example, \citeN{datta2008stencil} show that taking into account the overhead of data communication between CPU and GPU can significantly degrade GPU energy efficiency and can make them less energy efficient than CPUs. 

Similarly, some authors have found FPGAs to be more energy-efficient than GPUs \cite{kestur2010blas,hefenbrock2010accelerating,baker2007matched,thomas2009comparison,pauwels2012comparison,birk2012comparison,hussain2011highly,hamada2009comparative,gohringer2011reconfigurable,zou2012optimization,benkrid2012high,de2011energy,lange2009acceleration,williams2008computational,richardson2010comparative,leepower,van2012accelerating,brodtkorb2010state,thomas2013gpufpga,cong2009fpga,llamocca2011separable,cong20113d,waidyasooriyalow,chow2012mixed,wang2012coarse,lars2014gpufpga}, while others have found GPUs to be more energy efficient \cite{duan2011floating}. Similarly, some researchers observe other computing systems such as Cell, DSP (digital signal processor) or ASIC to be more energy efficient than GPUs \cite{chung2010single,baker2007matched,benkrid2012high,gpuDSP2011,pedram2012co}.

From these works, it is clear that although for majority of works, FPGAs are more energy efficient than GPUs and GPUs, in turn, are more energy efficient than CPUs, a single platform cannot be accepted as most energy efficient for all possible applications. The results crucially depend on the devices  and evaluation methodology used in the experiments.

\subsection{Discussion}

\citeN{keckler2011gpus} discuss the level of energy efficiency required for building future exascale machines. They show that for building an exascale machine with a power budget of 20MW requires an energy efficiency of 20 picoJoules (pJ) per floating point operation. In contrast, state-of-the-art CPUs and GPUs incur 1700 pJ and 225 pJ, respectively for each floating point operation. This shows that although the GPUs are more energy efficient than CPUs, their efficiency needs to be improved further to fulfill exascale challenge.      

\citeN{chandramowlishwaran2010optimizing} compare the performance and energy efficiency of a GPU with a multi-core CPU  for fast multipole method. They have observed that on applying suitable optimization and parallelization, the CPU is nearly 1.7$\times$ faster than a single GPU and achieves 0.75$\times$ the performance of two GPUs. In terms of energy efficiency, the CPU is nearly 2.4$\times$ and 1.8$\times$ as energy-efficient as the systems accelerated using one or two GPUs, respectively.

\citeN{datta2008stencil} compare the performance and energy efficiency of a GPU with a CPU for stencil (nearest-neighbor) computations. They observe that while use of large number of cores gives significant performance and power advantage to GPU over the CPU; when it is used as an accelerator offload engine for applications that primarily run on the host CPU, the performance and energy efficiency are severely degraded due to limited CPU-GPU bandwidth and low reuse within GPU device memory. Since the GPU can access CPU memory only through PCI-express (PCIe) bus, for applications which require larger on-board memory than what is available on the GPU, the performance is significantly degraded.

 \citeN{huang2009energy} evaluate the energy efficiency and performance of a GPU for a scientific computing benchmark, namely GEM software which is used to compute
the electrostatic potential map of macromolecules in a water solution. The CPU code is parallelized using  Pthread (POSIX threads). They observe that although the GPU consumes significantly higher power than the CPU, the execution time of GPU version of code is much smaller  and hence, the EDP (energy-delay product) of the GPU implementation is orders of magnitude better than that of both serial and parallel version of CPU implementation. Moreover, using a single-precision code  improves the energy efficiency of GPU even more.

\citeN{mcintosh2012benchmarking} compare the energy efficiency of a GPU with that of a multi-core CPU for a molecular mechanics problem. They observe that of the different GPU implementations tested, the best implementation outperforms all CPU implementations in both performance and energy efficiency. Moreover, for the real world case where the data set become larger, the benefits of GPU become even larger.
 
 \citeN{kestur2010blas} compare the energy efficiency and performance of a GPU with that of a multi-core CPU and an FPGA, for double-precision floating point programs from Basic Linear Algebra Subroutines (BLAS) library. They have shown that the FPGA offers comparable performance to GPU while providing significantly better energy efficiency. Moreover, the multi-core CPU also provides better performance and energy efficiency than the GPU.

 \citeN{llamocca2011separable} compare a GPU and an FPGA for 2D FIR (finite-impulse response) filter program which has application in video processing. They observe that due to its higher frequency and ability to exploit  massive parallelization present in the algorithm, the GPU provides better performance than the FPGA. However, the FPGA consumes up to an order of magnitude less energy than the GPU.

 \citeN{baker2007matched} compare the energy efficiency and performance of matched filter on an FPGA, an IBM Cell, a GPU and a CPU. Matched filter is a signal processing kernel which is used for extracting useful data from hyperspectral imagery. Relative to the CPU, the speedup of other computing systems is calculated and then a comparison is made on the metric of speedup and speed up per kilo Watt values. The authors observe that both Cell and FPGA outperform the GPU in performance and energy efficiency. Further, the GPU provides better performance and energy efficiency than the CPU.

 \citeN{hefenbrock2010accelerating} implement Viola-Jones face detection algorithm using multi-GPU and compare its performance and power consumption with that of fastest known FPGA implementation of the same algorithm. They observe that using 4-GPUs provides comparable performance with the design using a single FPGA, while the energy efficiency of FPGA design was orders of magnitude better than that of the 4-GPUs based design.

 \citeN{lange2009acceleration} compare the performance and energy efficiency of a GPU with an FPGA and a multi-core CPU for geometric algebra computations. They observe that the GPU is less energy efficient than the FPGA, but more efficient than the CPU. They also note that taking data transfer overhead into account degrades the energy efficiency of the GPU.

\citeN{hussain2011highly} compare the energy efficiency of a GPU with that of an FPGA and a CPU for k-means clustering algorithm, which is used in data mining. They observe that the FPGA provide better performance and energy efficiency than the GPU. Also, the GPU shows much better energy efficiency than the CPU.

\citeN{de2011energy} compare the energy efficiency of a GPU and a multi-core CPU with that of a hybrid FPGA-CPU implementation, for Monte Carlo option pricing with the Heston model. This program finds applications in financial domains. The hybrid FPGA-CPU implementation divides the work between FPGA and CPU, such that computation-intensive kernels are executed on FPGAs. They observe that compared to the GPU implementation, the hybrid FPGA-CPU implementation provides less performance but higher energy efficiency. Moreover, the GPU implementation excels CPU in both performance and energy efficiency.

\citeN{thomas2009comparison} compare energy efficiency of a GPU with an FPGA and a multi-core CPU for random number generation. The authors experiment with different random number generation programs and compute geometric mean of energy efficiency (number of samples generated per joule of energy). They observe that FPGAs provide an order of magnitude better energy efficiency than the GPU. Moreover, the GPU is found to be an order of magnitude more energy efficient than the CPU.

\citeN{van2012accelerating} implement random forest classification problem used in machine learning on a GPU, an FPGA and a multi-core CPU. They observe that the FPGA provides highest performance, but requires multi-board system even for modest size problems which increases its cost. Further, on performance per watt metric, the FPGA implementation is an order of magnitude better than the GPU implementation, which, in turn, is better than the CPU implementation.

\citeN{duan2011floating} compare a GPU with an FPGA and a multi-core CPU on floating-point FFT implementation. For GPU and CPU implementation they use standard libraries and for FPGA they develop their own implementation. They observe that the GPU is more energy efficient than the FPGA and the CPU for radix-2 FFT. They, however, observe a degradation in performance of the GPU for mixed-radix FFT.
     
\citeN{hamada2009comparative} make a comparative study of a GPU, an FPGA, an ASIC and a CPU for gravitational force calculation in $N$-body simulation in the context of astrophysics. They observe that the GPU outperforms the ASIC and the CPU in energy efficiency (performance per watt); however, its energy efficiency is an order of magnitude less than that of the FPGA.
 
\citeN{birk2012comparison} compare the performance and energy efficiency of a GPU and an FPGA for 3D ultrasound computer tomography which is used for medical imaging.  They observe that the performance of the GPU is comparable with that of the FPGA, however, the FPGA offer much better energy efficiency. 
  
 \citeN{betkaoui2010comparing} compare the energy efficiency of a GPU with an FPGA and a single and a multi-core CPU for three throughput computing applications, viz. FFT, general (dense) matrix multiplication (GEMM) and Monte Carlo method (MCM). Of these, GEMM is limited by computations, FFT by memory latency and MCM is embarrassingly parallel and hence is limited only by available parallelism. They use standard libraries for  implementing these applications. They observe that for all the three applications, the GPU outperforms the CPU on energy efficiency. Further, for GEMM, the GPU is more energy efficient than the FPGA, while for  FFT and MCM, the FPGA is more energy efficient than the GPU. They note that the FPGA provides advantage over the GPU for applications which exhibit poor data locality and low memory bandwidth requirement.

\citeN{zou2012optimization} compare a GPU with a CPU and an FPGA for Smith-Waterman (S-W) algorithm. S-W algorithm is used for performing pair-wise local sequence alignment in the field of bioinformatics. They highlight the need of making suitable optimizations on \textit{all} the three platforms for making meaningful comparisons. They observe that on the metric of performance per unit power, the FPGA is more energy efficient than the GPU, which in turn is more energy efficient than the CPU; although the advantage of GPU over CPU is small. The FPGA also provides higher performance than both GPU and CPU.

\citeN{benkrid2012high} compare a GPU with a CPU, an FPGA and Cell BE (broadband engine) for Smith-Waterman algorithm. They observe that on energy efficiency  (performance per watt) metric, the FPGA and Cell BE perform better than the GPU, while the GPU performs better than the CPU. They further note that results also depend on the devices used and performance optimizations performed on each platform.

\citeN{pauwels2012comparison} compare a GPU with an FPGA for the computation of phase-based optical flow, stereo, and local image features which is used in computer vision. They observe that while the GPU offers better performance and accuracy than the FPGA, the FPGA is  more energy efficient than the GPU.

 \citeN{fowers2013performance} compare the energy efficiency of a GPU with that of an FPGA and a multi-core CPU for convolution problem which has applications in digital signal processing. They observe that for very small signal sizes, the CPU is most energy efficient. However as the signal size increases, the energy efficiency of the GPU and the FPGA increase and for very large signal sizes, the FPGA outperforms GPU in energy efficiency.   

 \citeN{gpuDSP2011} implement high performance embedded computing (HPEC) benchmark suite on a GPU and compare the performance and energy efficiency of the GPU with that of a DSP for this benchmark suite. This benchmark includes a broad range of signal processing applications. They have observed that while the GPU provides at least an order of magnitude better performance than the DSP, its energy efficiency measured in terms of performance per watt is inferior to that of the DSP.

\section{Techniques for Improving GPU Energy Efficiency}\label{sec:improvement}
In this section, we discuss techniques for improving GPU energy efficiency. 

\subsection{Overview}
For the purpose of this study, we classify the techniques into the following categories. 
\begin{enumerate}
\item DVFS (dynamic voltage/frequency scaling) based techniques  \cite{liu2011waterfall,nam2007low,jiao2010power,lee2007dynamic,6337630,6270749,lee2011improving,sheaffer2005studying,chang2008energy,wang2010kernel,Liu2012PTM,ren2011algorithm,anzt2011analysis,6136741,lin2011power,zhao2012energy,huo2012energy,keller2010one,Abe2012PPA,park2006low,leng2013gpuwattch,paul2013coordinated}

\item CPU-GPU workload division based techniques \cite{takizawa2008sprat,rofouei2008energy,6337630,luk2009qilin,Liu2012PTM,liu2011waterfall,hamano2009power} and GPU workload consolidation \cite{li2011energy}

\item Architectural techniques for saving energy in specific GPU components, such as caches \cite{wang6176483,lee2011improving,lashgar2013inter,arnau2012boosting,Roger2013MICRO,lee2012tap}, global memory \cite{wang2013can,rhu2013locality}, pixel shader \cite{pool2011precision}, vertex shader \cite{pool2008energy}, core data-path, registers, pipeline and thread-scheduling \cite{abdel2013warped,6104522,gebhart2011energy,gilani2013exploiting,jing2013energy,yu2011sram,abdelwarped,gilani2012power,sethia2013pact}.  

\item Techniques which exploit workload-variation to dynamically allocate resources  \cite{jararweh2012power,liu2011waterfall,lee2011improving,hong2010integrated,6280275,6270749,6036835,keller2010one}

\item Application-specific and programming-level techniques for power analysis and management  \cite{6280275,chandramowlishwaran2010optimizing,suda2009power,datta2008stencil,jiao2010power,zandevakili2012gpumotif,anzt2011analysis,6136741,padoin2012evaluating,wang2010analysis,ghosh2012energy,dressler2012energy,6216757,wang2010kernel,yang2012fixing,hsiao2013energy} 

\end{enumerate}

We now discuss these techniques in detail. As seen through the above classification, several techniques can be classified in more than one groups. For sake of clarity, we discuss them in one group only.
 
\subsection{DVFS Based Techniques}
Dynamic voltage and frequency scaling (DVFS) is a well-known power-management technique which works by dynamically adjusting the clock frequency of a processor to allow  a corresponding reduction in the supply voltage to achieve power
saving. The relation between power and frequency is captured by the following formula \cite{rabaey2002digital}: 
\begin{equation}
 P \propto FV^2
\end{equation}
Here $F$ shows the operating frequency and $V$ shows the supply voltage. By intelligently reducing the frequency, the voltage at which the circuit needs to be operated for stable operation can also be reduced, leading to power saving. However, since the reduction in frequency also harms the performance, the scaling of voltage/frequency needs to be carefully performed.  Also note that in some of the works discussed below, the frequency scaling is actually applied to CPU; however, we still include these works since the power saving is achieved in the entire system and power management of CPU is done while taking into account the properties of GPU.

\citeN{nam2007low} propose a low-power GPU for hand-held devices. The proposed GPU uses logarithmic arithmetic to optimize area and power consumption. The use of logarithmic arithmetic leads to some computation error, however, due to the small screen of the hand-held devices, the error can be tolerated. They divide the chip into three power domains, viz. vertex shader, rendering engine and RISC processor, and DVFS is individually applied to each of the three domains.
The power management unit decides the supply voltage and frequency of each domain based on its workload for saving power while maintaining the desired performance level.  

\citeN{6136741} discuss an approach for saving system energy in a heterogeneous CPU-GPU computing system. They suggest that, instead of using a single GPU with each CPU, using multiple GPUs with each CPU enables achieving speedup in execution time and improving the usage of CPU, which improves the energy efficiency of the system. Further, since during the execution of CUDA kernel, the host CPU remains in polling loop without doing useful work, the frequency of CPU can be reduced for saving energy while always ensuring that CPU frequency is greater than the PCIe bus between CPU and GPU. Since the range of high-performance CPU frequencies are generally larger than that of PCIe bus, CPU frequency can be scaled without affecting GPU performance. They demonstrate their approach by parallelizing 3-D finite element mesh refinement on GPU. 
  
\citeN{anzt2011analysis} propose techniques for reducing energy consumption in CPU-GPU heterogeneous systems for executing iterative linear solvers. They propose using DVFS for saving energy in CPU while it stays in busy-wait waiting for GPU to complete computations. Since during this time, CPU performs no useful work, use of DVFS gives large energy saving with little performance loss. Further, since the conjugate gradient iterative linear solver consumes nearly same time in different iterations; by noting this duration once, the CPU can be transitioned to sleep state for this duration in further calls to the kernel; which leads to further energy savings. They also remark that use of this technique is useful when the calls to kernels consume a sufficiently large amount of time.

\citeN{jiao2010power} study  the the performance and power consumption of GPU for three computationally diverse applications
 for varying processor and memory frequencies. Specifically, they study dense matrix multiplication (compute-intensive), dense matrix transpose (memory-intensive), and fast Fourier transform (hybrid). They have observed that the power consumption of GPUs is primarily dependent on the ratio of global memory transactions to computation instructions  and the rate of issuing instructions. These two metrics  decide whether an application is memory-intensive or computation-intensive, respectively. Based on these characteristics, the frequency of GPU cores and memory is adjusted to save energy.

   \citeN{lin2011power} propose use of software-prefetching and dynamic voltage scaling to save GPU energy. Software-prefetching is a technique which aims to improve performance by overlapping the computing and memory access latencies. It works by inserting prefetch instructions into the program so that data is fetched into registers or caches well-before time and processor-stall on memory access instructions is avoided. Since prefetching increases the number of instructions, it also increases the power consumption and hence, it must be balanced with suitable performance enhancement. Their technique analyzes the program to insert prefetching instructions and then iteratively uses DVFS to find a suitable frequency such that performance constraint is met while saving largest possible amount of energy.

\subsection{CPU-GPU Work Division to to Improve Energy Efficiency}
Researchers have shown that different ratios of work-division between CPUs and GPUs may lead to different performance and energy efficiency levels \cite{6337630,luk2009qilin}. Based on this observation, several techniques have been proposed which dynamically choose between CPU and GPU as a platform of execution of a kernel based on the expected energy efficiency on those platforms.

\citeN{6337630} propose an energy-management framework for GPU-CPU heterogeneous architectures. Their technique works in two steps. In the first step, the workload is divided between CPU and GPU based on the workload characteristics, in a manner that both sides may complete their tasks approximately at the same time. As an example, the task shared of CPU and GPU may be 15\% and 85\%, respectively. This step ensures load-balancing which also avoids the energy-waste due to idling. In the second step, the frequency of GPU cores and memory are adjusted, along with the frequency and voltage of the CPU to achieve largest possible energy savings with minimal performance degradation.

\citeN{luk2009qilin} propose an automatic technique for mapping computations of processing elements on a CPU/GPU heterogeneous system. Compared to other approaches which require the programmer to manually perform the computations to processor mapping, their technique uses run-time adaptation to  automatically perform the mapping.  Their technique provides an API (application programming interface) for writing parallelizable programs. Through the API, the computations are explicitly expressed and hence, the compiler is not required to extract parallelism from the serial code. While OpenMP can exploit parallelism only on CPU, their technique can exploit parallelism on both the CPU and the GPU. Since the optimal mapping changes with different applications, hardware/software configurations and input problem sizes, the adaptive mapping outperforms hand-tuned mapping in both performance and energy efficiency.

\citeN{Liu2012PTM} discuss a technique for finding power-efficient mappings of time-critical applications onto CPU/GPU heterogeneous systems. Their technique works in two steps. In the first step, their technique maps the application to either CPU or GPU, such that their deadlines are met and execution time is minimized. In the second step, DVFS techniques are applied to both CPU and GPU to save energy. The mapping of applications can be done in both offline and online manner. To keep the  performance high and avoid resource-idling, their technique also aims to achieve load-balancing. Moreover,  their technique utilizes the fact that typically average-case execution times are less than their worst-case execution time and hence, early completion provides a slack which can be exploited using DVFS to save large amount of energy.

\citeN{takizawa2008sprat} propose SPRAT (stream programming
with runtime auto-tuning), a runtime environment for dynamically selecting a CPU or GPU with a view to  improve the energy efficiency. They introduce a performance model which takes into account the relative execution time and energy consumption on CPU and GPU and the data transfer time between CPU and GPU. This model is especially suited for applications that require frequent data transfers between CPU and GPU. Based on the runtime behavior,  SPRAT can dynamically select the computing platform (CPU or GPU) for executing a kernel such that system energy is minimized.

\citeN{rofouei2008energy} experimentally evaluate the power and energy cost of GPU operations and compare it that of CPU for convolution problem. They find the relation between execution time and energy consumption and show that that GPU is more energy efficient when it provides application performance improvement above a certain threshold. Based on this, the decision about running the application on CPU or GPU can be taken.  \citeN{suda2009power} discuss a scenario where the performance benefit provided by using two GPUs (instead of one) offsets the power consumption overhead of the extra GPU and leads to power saving. They demonstrate their approach for  multiplication of large matrices.   

 \citeN{liu2011waterfall} develop an energy saving algorithm for large scale GPU cluster systems based on the waterfall model. In their cluster, each node may have many CPU-GPU pairs.  Their method divides the energy consumption of overall system into three different levels based on different energy saving strategies deployed. Their method formulates the energy saving problem as an optimization task, where the energy consumption needs to be minimized while meeting task deadlines. Their technique transitions the node in one among three states, namely busy (all CPUs and GPUs inside a node are executing task), spare (at least one CPU-GPU pair is free) and sleep (all CPU-GPU pairs are free). At the time of reduced workload, the node in sleep state is powered off to save energy and at time of additional workload, a node is woken up. Also, their technique selects an appropriate task from the set of available tasks and schedules it on optimal CPU-GPU pair such that the execution time of the task is minimized. Further, the voltage of the CPU is adaptively scaled to save energy while meeting task deadline. Finally, they also utilize $\beta$-migration policy, where a small fraction ($\beta$) of the GPU's share of task is migrated to the CPU in the same CPU-GPU pair for achieving load-balancing.

\subsection{Saving Energy in GPU components}
Several techniques make architecture-level changes to GPUs to optimize the energy spent in individual components of the GPU. These techniques utilize the specific usage pattern of GPU components to make runtime adaptation for saving energy.       

\citeN{gebhart2011energy} present a technique for saving energy in core datapath of GPU. Since GPUs employ a large number of threads, storing the register context of these threads requires a large amount of on-chip storage. Also, the thread-scheduler in GPU needs to select a thread to execute from a large number of threads. For these reasons, accessing large register files and scheduling among a large number of threads consumes substantial amount of energy. To address this, Gebhart et al. present two improvements. First, a small storage structure is added to register files which acts like a cache and  captures the working set of registers to reduce energy consumption. Second, the threads are logically divided into two types, namely, active threads (which are currently issuing instructions or waiting on 
relatively short latency operations), and pending threads (which are waiting on long memory latencies). Thus, in any cycle, the scheduler needs to consider only the active threads which are much smaller in number. This leads to significant energy savings.

\citeN{wang6176483} propose a technique for saving static energy in both L1 and L2 caches. They propose putting L1 cache (which is private to each core) in state-preserving\footnote{State-preserving  refers to the low-power state where the contents stored in the block are not lost. This is in contrast with state-destroying low-power state where the block contents are lost in the low-power mode \cite{masterMittal2013}.} low-leakage mode when there are no threads that are ready to be scheduled. Further, L2 cache is transitioned to low-leakage mode when there is no memory request. They also discuss the microarchitectural optimizations which ensure that the latency of detecting cache inactivity and transitioning a cache to low-power and back to normal power are completely hidden.

\citeN{lashgar2013inter} propose the use of filter-cache to save energy in GPUs by reducing accesses to instruction cache. Their technique is based on ``inter-warp instruction temporal locality'' which means that during short execution intervals, a small number of static instructions account for a
significant portion of dynamic instructions fetched and decoded within the same stream multiprocessor. Thus, the probability that a recently fetched instruction will be fetched again is high. They propose using a small filter-cache to cache these instructions and reduce the number of accesses to instruction cache, which improves the energy efficiency of the fetch engine. Filter-cache has been used in CPUs also, however, in GPUs the instruction temporal locality is even higher. This is because GPUs interleave thousands of threads per core, which are grouped in warps. The warp scheduler continuously issues instructions from different warps which fills the warp, thus fetching the same instruction for all warps during short intervals.

A unified local memory design for  GPUs is presented by \citeN{gebhart2012unifying}. The existing GPUs use rigid partition sizes of registers, cache, and scratchpad, however, different GPU workloads have different requirements of registers, caches and scratchpad (also called shared memory).  Based on the characterization study of different workloads, they observe that different kernels and applications have different requirements of cache, shared memory etc. To address this issue, they propose a unified memory architecture that aggregates these three types of
storage and allows for a flexible allocation on a per-kernel basis. Before the launch of each kernel, the system reconfigures the memory banks to change the  partitioning of the memory. By effectively using the local-storage, their design reduces the accesses to main memory. They have shown that using their approach broadens the range of applications that can be efficiently executed on GPUs and also provides improved performance and energy efficiency.

To filter a large fraction of memory requests that are serviced by the first level cache or scratchpad memory, 
 \citeN{sankaranarayanan2013Tiny} propose adding small sized caches (termed as tinyCaches) between each lane in a streaming multiprocessor (SM) and the L1 data cache which is shared by all the lanes in an SM.  Further, using some unique features of CUDA/OpenCL programming model, these tinyCaches avoid the need of complex coherence schemes and thus, they can be implemented with low-cost.  They have shown that their design leads to improvement in the energy efficiency of the GPU.

\citeN{rhu2013locality} propose a technique for finding the right data-fetch granularity for improving performance and energy-efficiency of GPUs. They observe that only few applications use all the four 32B sectors of the 128B cache-block, which leads to over-fetching of data from the memory. To address this issue,  their technique first decides the appropriate granularity (coarse-grain or fine-grain) of data fetch.  Based on this, a hardware predictor adaptively adjusts the memory access granularity without programmer or runtime system intervention. Thus, their approach enables adaptively adjusting the memory access granularity depending on the spatial locality present in the application.

In a CPU-GPU heterogeneous computing system (HCS) with shared last level cache (LLC), interference between CPU and GPU threads can lead to degradation in performance and energy efficiency. This is especially critical since the GPU has much larger number of threads than the CPU,  and  hence, the large number of accesses from GPU are likely to evict data brought in cache by the CPU threads. Some authors propose techniques to address this issue \cite{lee2012tap,mekkat2013managing}. \citeN{lee2012tap} propose a thread-level parallelism (TLP) aware cache management policy for such systems.  Due to the presence of  deep-multithreading, a cache policy does not directly affect the performance in GPUs. Hence, to estimate the effect of cache behavior on GPU performance, they propose a core-sampling approach, which leverages the fact that most GPU applications show symmetric behavior across the running cores. Based on this, core sampling applies a different policy (e.g. a cache replacement policy) to each core and periodically collects samples to see how the policies work. A large difference in performance of these cores indicates that GPU performance is affected by the cache policy and vice versa. Using this, the best cache management policy can be chosen. Further, to alleviate the interference,  they introduce cache block lifetime normalization approach, which ensures that statistics collected for each application are normalized by the access rate of each application. Using this, along with a cache partitioning mechanism, cache is partitioned between CPU and GPU, such that cache is allocated to GPU only if it benefits from the cache.

\citeN{mekkat2013managing} propose a technique which leverages GPU's ability to tolerate memory access latency to throttle GPU LLC accesses to provide cache space to latency-sensitive CPU applications. Based on the observation that the TLP available in an application is a good indicator of cache sensitivity of an application,  their technique allows GPU memory traffic to selectively bypass the shared LLC if GPU cores exhibit sufficient TLP to tolerate memory access latency or when GPU is not sensitive to LLC performance. A large number of wavefronts that are ready to be scheduled indicates a higher amount of TLP. Using core-sampling, they apply two different bypassing thresholds to two different cores to find the impact of bypassing on GPU performance. Also, using cache set-sampling, the effect of GPU bypassing on CPU performance is estimated. Using these, the rate of GPU bypassing is periodically adjusted to improve performance and save energy.

\subsection{Dynamic Resource Allocation Based Techniques}
It is well-known that there exists large intra-application and inter-application variation in the resource requirements of different applications. In fact, several real-world applications rarely utilize all the computational capabilities of GPU. Thus, significant amount of energy saving can be achieved by dynamically adapting the components which exhibit low utilization levels. 

\citeN{hong2010integrated} propose an integrated power and performance prediction system to save energy in GPUs. For a given GPU kernel, their method predicts both performance and power; and then uses these predictions to choose the optimal number of cores which can lead to highest performance per watt value. Based on this, only desired number of cores can be activated, while the remaining cores can be turned-off using power-gating. Note that power-gating is a circuit-level scheme to remove leakage by shutting-off the supply voltage to unused circuits.

\citeN{wang2011power} propose power-gating strategies for saving energy in GPUs. In graphics applications, different scenes have different complexities (e.g. number of objects) and hence, the amount of computing resources which are required to provide a satisfactory visual perception varies across different frames. By predicting the required shader resources for providing desired frame-rate, the extra shader resources can be turned-off using power-gating. To avoid the overhead of power-gating, their technique ensures that the idle period of the unused circuits is long enough to compensate the switching overhead.

\citeN{6036835} present an offline profiling based technique to estimate the appropriate number of GPU cores for a given application to save energy.  Their technique uses the profile of PTX (parallel thread execution) codes generated during compilation of the application to decide the number of cores to be used for achieving highest energy efficiency. During actual run, in place of using programmer-specified number of cores, only the desired number of cores can be activated to save energy.

Among the commercial products, AMD uses PowerPlay technology \cite{amdpowerplay} for dynamic power management. It dynamically transitions the GPU between low, medium and high states, based on the load on the GPU. For example, while a graphics application is running, the demand on GPU is high and hence it runs in high power state. Conversely, while typing emails, the load on GPU is minimal and hence, it runs in low power state. The power saving also reduces system temperatures and the fan noise. Similarly, NVIDIA uses PowerMizer technology for dynamic power management \cite{nvidiapowermizer}.

\subsection{Application-specific and programming-level techniques}
It has been observed that source-code level transformations and application-specific optimizations can significantly improve the resource-utilization, performance and energy efficiency of GPUs. Thus, by performing manually or automatically optimizing GPU implementation and addressing performance bottlenecks, large energy savings can be obtained.      

\citeN{wang2010kernel} propose a method for saving energy in GPU using kernel-fusion. Kernel fusion combines the computation of two kernels into a single thread.  Thus, it leads to balancing the demand of hardware resources, which improves utilization of resources and thus, improves the energy efficiency. The authors formulate the task of kernel-fusion as a dynamic programming problem, which can be solved using conventional tools.

\citeN{6280275} propose a technique to save energy in task-parallel execution of dense linear algebra operations (viz. Cholesky and LU factorization),  by intelligently replacing the busy-waits with a  power-friendly blocking state. Execution of these tasks involves CPU thread issuing the kernel (for execution on GPU) and then waiting for the next ready task in a busy-wait polling loop. This leads to wastage of energy. To avoid this, their technique blocks the CPU thread on a synchronization primitive when waiting for the GPU to finish work; thus leading to saving of energy.

\citeN{ghosh2012energy} study the energy efficiency of HPC application kernels (viz. matrix-matrix multiplication, FFT, pseudo-random number generation and 3D finite difference) on multi-GPU and multicore CPU platforms. The kernel implementations are taken from standard libraries. They observe that while the absolute `power' consumption (in Watts) of multi-GPU is larger than that of the multicore CPU, the `energy efficiency' (in Giga Flops per Watt ) of GPUs is much superior than that of CPUs. They observe that for GPUs, the number of global memory accesses and operations per unit time have significant influence on the power consumption. Also, a large computation to communication ratio per device is important for hiding data transfer latency and realizing energy efficiency in GPUs. 

\citeN{yang2012fixing} evaluate several open-source GPU projects  and suggest ways to change the program code to improve GPU usage, performance and energy efficiency. These projects are taken from a wide range of disciplines, such as atmosphere science,
computational physics, machine learning, bioinformatics and mathematics. They identify the common code patterns which lead to inefficient hardware use. For example, adjustment of thread-block dimension can improve the way global memory data are  accessed and reused in either shared memory or hardware caches.  Further, choice of global memory data types and use of texture and constant memory has significant effect on achieved bandwidth. Also, by optimizing the program for specific GPU (e.g. AMD GPU or NVIDIA GPU), the hardware-specific features can be exploited to obtain higher performance and energy efficiency.

\section{Future Research Trends and Conclusion}\label{sec:conclusion}

We believe that in the near future, the challenges of GPU power consumption would need to be simultaneously addressed at different levels at the same time. At the chip-design level, researchers are aiming to develop energy-efficient throughput cores and memory design to exploit instruction-level, data-level and fine-grained task-level  parallelism. At the architecture level, CPU and GPU need to be efficiently integrated on the same chip with a unified memory architecture \cite{foley2012low,yuffe2011fully}. This will address the memory bandwidth bottleneck and also avoid the replicated chip infrastructure and the need of managing separate memory spaces. At the programming level, per-application tuning is inevitable to achieve a fine balance between demands of the application and the resources of the GPU. Finally, at the system level, policies for intelligent scheduling and work-division between CPU and GPU are required, so that their individual competencies are integrated and they complement each other. 

The 3D die stacking holds the promise of mitigating memory bandwidth bottleneck in GPUs, as it enables use of shorter, high-bandwidth and power-efficient global interconnect and provides denser form factor. 3D stacking also enables integration of heterogeneous technologies, which allows use of non-volatile memory (NVM), such as phase change RAM (PCM) and spin transfer torque RAM (STT-RAM) in the design of GPU memory \cite{mittal2013PCM}. NVMs consume negligible leakage power  and provide higher density than SRAM and DRAM, however, their write latency and energy are significantly higher than those of SRAM and DRAM. It is expected that leveraging the benefits of 3D stacking and NVM would be a major step in improving the energy efficiency of  GPUs and it would require novel solutions at device, architecture and system level.

As GPUs become deployed in large-scale data-centers and supercomputers, the challenges of power management are expected to grow. For such large systems, power management needs to be done at the level of both intra-node and inter-node. These nodes may be remotely situated and may have different configurations (e.g. CPU, GPU, FPGA etc. or different interconnection). Managing power consumption of such systems while taking into account load-balancing, temperature reduction and performance-target will be an interesting research problem for the designers. On the other side of the spectrum, in battery-operated devices such as smartphones, where the need of processing visually compelling graphics within a small power budget increases  with each new generation, the requirement for aggressive energy optimization will pose novel challenges for the computer architects.

Virtualization technology enables multiple computing environments to be consolidated in a single physical machine and thus, increases resource utilization efficiency and reduces total cost of ownership (TCO). Specifically, in cloud computing, virtualization is key enabling technology since flexible resource provisioning is essential for unpredictable user demands. Very recently, GPUs have been used in cloud-computing and virtual-machine (VM) platforms \cite{nvidiaGPUVM,jo2013exploiting,shi2012vcuda,amazoncloud}.   By adding or removing GPUs in each VM in on-demand manner, VMs in the same physical host can use the GPUs in time-sharing manner \cite{jo2013exploiting}, which also leads to significant reduction in idle power of GPUs. We believe that much research still needs to be done to leverage virtualization for minimizing power and TCO of GPU computing infrastructure.

In this paper, we surveyed several methods aimed at analyzing and improving the energy efficiency of GPUs. We underscored the need of power management in GPUs and identified important trends which are worthy of future investigation. Further, we presented a classification of different research works to highlight the underlying similarities and differences between them. We believe that this survey will provide the researchers valuable insights into the state-of-the-art in GPU power management techniques and motivate them to create breakthrough inventions for designing green GPUs of the future exascale era.

\DeclareRobustCommand{\disambiguate}[3]{#3}
\bibliographystyle{ACM-Reference-Format-Journals}
\bibliography{References}


\end{document}